\documentclass[aps,pra,preprint,superscriptaddress]{revtex4}

\usepackage{graphicx}
\usepackage{amssymb}
\small
\begin{document}
\title{Pull-in control in microswitches using acoustic Casimir forces }

\author{R. Esquivel-Sirvent$^{1}$ and L. Reyes  }
\email[Corresponding author. Email:]{raul@fisica.unam.mx}
\affiliation{Instituto de F\'{\i}sica, Universidad Nacional Aut\'onoma
 de M\'exico, Apartado Postal 20-364, D.F. 01000,  M\'exico}

\begin{abstract}
In this paper we present a theoretical calculation of the acoustic Casimir pressure in a model micro system.  Unlike the quantum case, the acoustic Casimir pressure can be made attractive or repulsive depending on the frequency bandwidth of the acoustic noise. 
As a case study, a one degree of freedom simple-lumped system in an acoustic resonant cavity is considered. 
We show that the frequency bandwidth of the acoustic field  can be tuned to increase the stability in existing microswitch systems by selecting the sign of the force. The acoustic intensity and frequency bandwidth are introduced as two additional control parameters of the microswitch.
\end{abstract}

\maketitle

\section{Introduction} 
The term acoustic Casimir force (ACF) refers to the force between two parallel plates when they are placed in an acoustic random field. This is a classical analog of the quantum Casimir force that results from quantum vacuum fluctuations \cite{casimir,milonni}.
Acoustic Casimir forces were first proposed by Larraza and collaborators \cite{larraza,larraza2}, who measured the effect in audible frequencies for plates separated a few millimeters. The plates were placed in
a close tank that acted as a reverberation chamber, and the average  acoustic field  of intensity $I$  was generated with pressure drivers  in a frequency range $[\omega_1,\omega_2]$.  In the acoustic Casimir force,  only the frequencies in  a finite bandwidth spectrum are taken into account, the bandwidth being  determined by the physical limits of the  acoustic sources used.  Unlike the unbounded spectrum of the quantum case, the ACF has very interesting physical consequences.  The most significant being that the acoustic  Casimir force  changes from attractive to repulsive depending on the plate separation and the frequency bandwidth.  The change in sign of the force due to a finite bandwidth has also been observed in the fermionic Casimir effect  between two metals, where the force is obtained after an integration over all possible energy states up to the Fermi level \cite{villarreal}.

 In this paper,   an external classical sound source is assumed in the calculation  of the ACF. It differs from the phononic Casimir effect,  that is caused by  a random phonic field induced by thermal fluctuations  \cite{bschorr}. The  magnitude of the phononic Casimir pressure is comparable to the quantum Casimir effect  at separations of the order of $10^{-5}$ m.  The phononic pressure  has been shown to influence the pull-in dynamics of micro electromechanical systems (MEMS) \cite{george07}. 

\section{Acoustic Casimir Pressure}

Consider  two parallel plates separated a distance $L$ and characterized by their acoustic reflectivities $r$. The plates are in an acoustic field of random white noise of frequency bandwidth $[\omega_1,\omega_2]$ and spectral intensity \cite{footnote1} $I_{\omega}$. The acoustic Casimir pressure (ACP) is calculated from \cite{jasa}

\begin{equation}
P=-\frac{I_{\omega}}{\pi}\int_{\omega_1/c}^{\omega_2/c} dk_z \int_{\sqrt{\omega_1^2/c^2-k_z^2}}^{\sqrt{\omega_2^2/c^2-k_z^2}} dQ \frac{k_z^2 Q}{k^4}\left (\frac{1}{\xi-1} \right)
\label{lifshitz2}
\end{equation}
where $\xi=(r^2 exp(2ik_zL))^{-1}$. The wave vector $k^2=\omega^2/c^2=Q^2+k_z^2$ has components $Q$ parallel to the plates and $k_z$ perpendicular to the plates. The speed of sound in the medium between the plates is $c$. 
Equation (\ref{lifshitz2}) was derived \cite{jasa} assuming rigid plates, so that no sound is generated by them.  Also, diffraction effects are neglected.

The basic idea behind Eq (\ref{lifshitz2}) is the subtraction of the pressure outside the plates from the  pressure due the modes that can exists between the plates when the boundary conditions are satisfied.  If the lower frequency of the bandwidth is set to zero, the ACP is always attractive \cite{larraza}.    If we consider and infinite bandwidth $\omega \in [0,\infty)$ and perfect acoustic reflectors $r=1$, the ACP
calculated using Eq.(\ref{lifshitz2} ) gives a attractive force for all separations, 
\begin{equation}
P_0=-\frac{\pi I_{\omega}}{4L}.
\label{lifshitz3}
\end{equation}

To scale down the ACP from macroscopic systems  to submicron systems, we require high frequencies.  Typical commercial transducers can operate up to $250$ MHz, and with a high acoustic power output.  The feasibility of high intensity, high frequency transducers in micro systems can be exemplified by the experiments of Degertekin \cite{degertekin} that actuated atomic force microscopes cantilevers using focused ultrasonic transducers in the frequency range of 100-300 $MHz$ . Also, Sub-teraHertz have been achieved with superlattice structures \cite{huynh}. The  intensity of ultrasonic transducers is around 50 $mW/cm^2$ \cite{uchida}. However, high intensity focused ultrasonic transducers can have an intensity of several orders of magnitude greater; several thousands Watts per square centimeter \cite{zhou}. 

\section{Applications to micro switches}

To study the effect of the acoustic Casimir force in mems, 
 we consider a  one degree of freedom  simple lumped model.    It  consists of two parallel plates, one fixed the other attached to a spring of elastic constant $k$. The equilibrium separation is $D=60$$\mu m$. When the plates are at a distance $L$, the elastic force is $F_e=-k(D-L)$.  The system in enclosed in an acoustic resonant cavity as shown in Figure 1.  Although this model is the simplest when studying pull-in dynamics, it  is helpful in the overall understanding of the pull-in dynamics and in our case,  how it changes in the presence of the acoustic Casimir force. For these systems, the electrostatic force  is the most common form of actuation for MEMS.
To get an idea of the pressure magnitudes involved in the Acoustic Casimir effect, we compare the ACP  for perfect acoustic reflectors and an infinite bandwidth, Eq.(\ref{lifshitz3}), with the pressure between two parallel plates of area $A$ with a potential difference between them.  We consider two cases, when the potential difference is of $3$$V$  and of $6$$V$, as shown in Figure 2. In these cases, they are of the same order of magnitude.    Recall, that the ACP is proportional to the spectral intensity $I_{\omega}$, that  we can vary to  get a Casimir pressure  of the same order of magnitude as the electrostatic case.  In Figure 1 we chose $I_{\omega}=10^{-4}$$Watts\cdot s^{-1}\cdot m^{-2}$.

The important  aspect of the ACP we want to emphasize is that it may change sign when the plate separation changes.   As explained in Ref. \cite{larraza}, the repulsive Casimir pressure happens when the separation  between the plates equals an integral number of half-wavelengths associated with the lower frequency of the band-width  $\omega_1$.  In this case, the  wave vector $k\sim k_z$ between the plates and the contribution of the stress tensor is mainly that perpendicular to the plates.   Outside the plates this condition is not met and the for the same frequency the wave vector can be at any angle of incidence.  In Figure 3, we present the ACP as a function of separation for two different frequency band widths $[90,100]$ $MHz$ (solid line) and $[90,100]$ $GHz$  (dotted line).
The repulsive ACP  should occur when $L\sim \pi  n c /\omega_1$, that is, close to any of the following separations : 11.8682, 23.7365, 35.6047, 47.473, 59.3412, 71.2094, 83.0777,
94.9459, 106.814, 118.682, 130.551, 142.419.  For the MHz bandwidth these positions are in microns and for the GHz bandwidth in nanometers.  This shows that the ACP is applicable to devices of any size by a suitable choice of the frequency bandwidth.

The peaks where the repulsive ACP occurs get broader as the separation increases. At large separations, the system is less sensitive to variations in frequency. If $\Delta L$ be the peak width,  the change in the frequency within this interval is $\Delta \omega \sim n\pi c/\Delta L$.  Larger $\Delta L$ imply a small  change in the frequency $\Delta \omega$ as well as  in wave vector. Thus, within $\Delta L$ at large separations $k\sim k_z$ within the plates.

Controlling the lower frequency of the bandwidth allows us to select where the ACP changes sign.  Consider the ACP for the bandwidth 
$[25,28]$$MHz (GHz)$. This, is shown in Figure 3. There is a narrow peak centered at $L=40$ $\mu m (nm)$ or $L=2D/3$, where a repulsive ACP is present, while for other separations we have  a constant negative pressure. Again, changing the bandwidth from MHz to GHz allows us to work with different size systems. Let us recall that for the simple lumped system actuated by electrostatic forces \cite{batra,pelesko}   there is an upper limit for the value of the potential difference after which the electrostatic force overcomes the elastic force and the top plate jumps to contact.  The voltage $V_{in}$ and plate separation $L_{in}$ where the pull-in occurs are
 \begin{eqnarray}
 V_{in}&=&\sqrt{\frac{8kD^3}{27 \epsilon_0 A}}\\ \nonumber  
 L_{in}&=&\frac{2}{3}D
 \label{voltin}
 \end{eqnarray}
this  is why we choose the ACP to have a positive peak at this particular separation.  Now, we show that having the acoustic Casimir pressure, the dynamics of the plate changes and the pull-in voltage can be increased.  Having another force besides the electrostatic attraction has been considered in the study of pull-in dynamics,  for example dispersive forces \cite{delrio,zhao07,spinello07}. 

To find the critical value of the voltage where the pull-in occurs, we begin with the equation of motion of the top plate
\begin{equation}
F=-k(D-L)+\frac{\epsilon_0 V^2 A}{2 L^2}+F_{ac},
\label{motion}
\end{equation}
where   $\epsilon_0$ the permittivity of vacuum.   Introducing the dimensionless separation $\tilde{L}=D/L$  we have \cite{pelesko}
\begin{equation}
(1-\tilde{L})+\frac{\lambda_1}{\tilde{L}^2}+\lambda_2  \int d{\tilde Q} d{\tilde k}_z\frac{{\tilde k}_z^2 {\tilde Q}}{{\tilde k}^4}\left (\frac{1}{{\tilde \xi}-1} \right)=0.
\label{dimensionless}
\end{equation}
 The wavevector components with a tilde are normalized to $k_0=1/D$, ${\tilde \xi=(r_1r_2 exp(2i{\tilde k_z}{\tilde L}))^{-1}}$. In Eq.(\ref{dimensionless}) we have introduced the parameters
\begin{equation}
\lambda_1=\frac{\epsilon_0AV^2}{2 k D^3},
\label{lambda1}
\end{equation}
 which shows the relative importance of the electrostatic force to the elastic force, and
\begin{equation}
\lambda_2=\frac{I_{\omega}}{2\pi^2 k D^2}.
\label{lambda2}
\end{equation}
 that compares the acoustic Casimir force with the elastic force.
 Since the voltage appears in $\lambda_1$ we find that
 \begin{equation}
 \lambda_1={\tilde L}^2(1-{\tilde L})+\lambda_2 f(\tilde{L}),
 \label{lam1}
 \end{equation}
 where we define $f(\tilde{L})$ is the integral term that multiplies $\lambda_2$ in Eq. (\ref{dimensionless}).

 If $\lambda_2=0$, the solution of Eq. (\ref{lam1}) corresponds to the pull-in voltage given in Eq.(3). For the the value of $D=60$$\mu m$ used here yields $L_{in}=40$$\mu m$. In Fig(3), we choose the frequency bandwidth to have a positive pressure at this separation. As we now show, the additional ACP can increase the value of the pull-in voltage.  

 For different values of $\lambda_2$, we plot the bifurcation diagram   $\lambda_1$ vs  ${\tilde L}$ in Fig. (4). The vertical line is a visual aid that indicates the position of the maximum of the curve .  The points to the left of the maximum are stable equilibrium point while all the points to the right are unstable points and for those values of $\lambda_1$ and $\tilde {L}$ the elastic force is overcome and the plates will jump to contact \cite{zhao07}.  For the particular selection of the ACP used the position of the maximum does not change for different values of $\lambda_2$, since $d f(\tilde{L})/d {\tilde{L}}=0$, as seen in Fig. (4). So, for this particular choice of frequency the bandwidth is still $L_{in}=\frac{2}{3}D$. However, the pull-in voltage will change.  Using the definition of $\lambda_1$ in Eq. (\ref{lam1}) and evaluating at  $L_{in}$ we have
 \begin{equation}
 V^{*}=V_{in}[1+\frac{27}{4}\lambda_2 f(L_{in})]^{1/2},
 \end{equation}
  where $V_{in}$ is the pull-in voltage in the electrostatic case, Eq.(\ref{pullin}), since $\lambda_2$ is proportional to the acoustic intensity. Thus,  the pull-in voltage can be increased by increasing the acoustic intensity. 

\section{Conclusions}

In conclusion, we have presented a theoretical calculation of the acoustic Casimir force in frequency ranges suitable for applications in the micrometer range.  In particular, we showed that by selecting a particular frequency bandwidth the ACP can be repulsive or attractive, or change sign as a function of plate separation.  The lower frequency of the bandwidth and the acoustic intensity are two parameters that can be tuned to control the dynamics of electrostatic actuated MEMS extending the value of the pull-in voltage for simple-lumped systems. The  ACP calculations were done for a micrometer size and nanometer size systems.  However, these Therefore, a fine-tuned ACP can be used to increase the mechanical stability of MEMS structures. In future work, by considering oscillating ACP, as shown in Fig. (2), we will analyze the possibility of using the Acoustic Casimir effect as a mechanical actuator for micropumps \cite{nguyen}. Another area of interest where the analogy between the quantum case and the acoustic can be exploited is in the use of acoustic metamaterials \cite{metaac}, where negative volume densities and reflectivities are possible.

\acknowledgements{Partial support for this project was provided by DGAPA-UNAM Grant No. IN113208}.

\newpage

\begin{figure}
\includegraphics[width=8.cm]{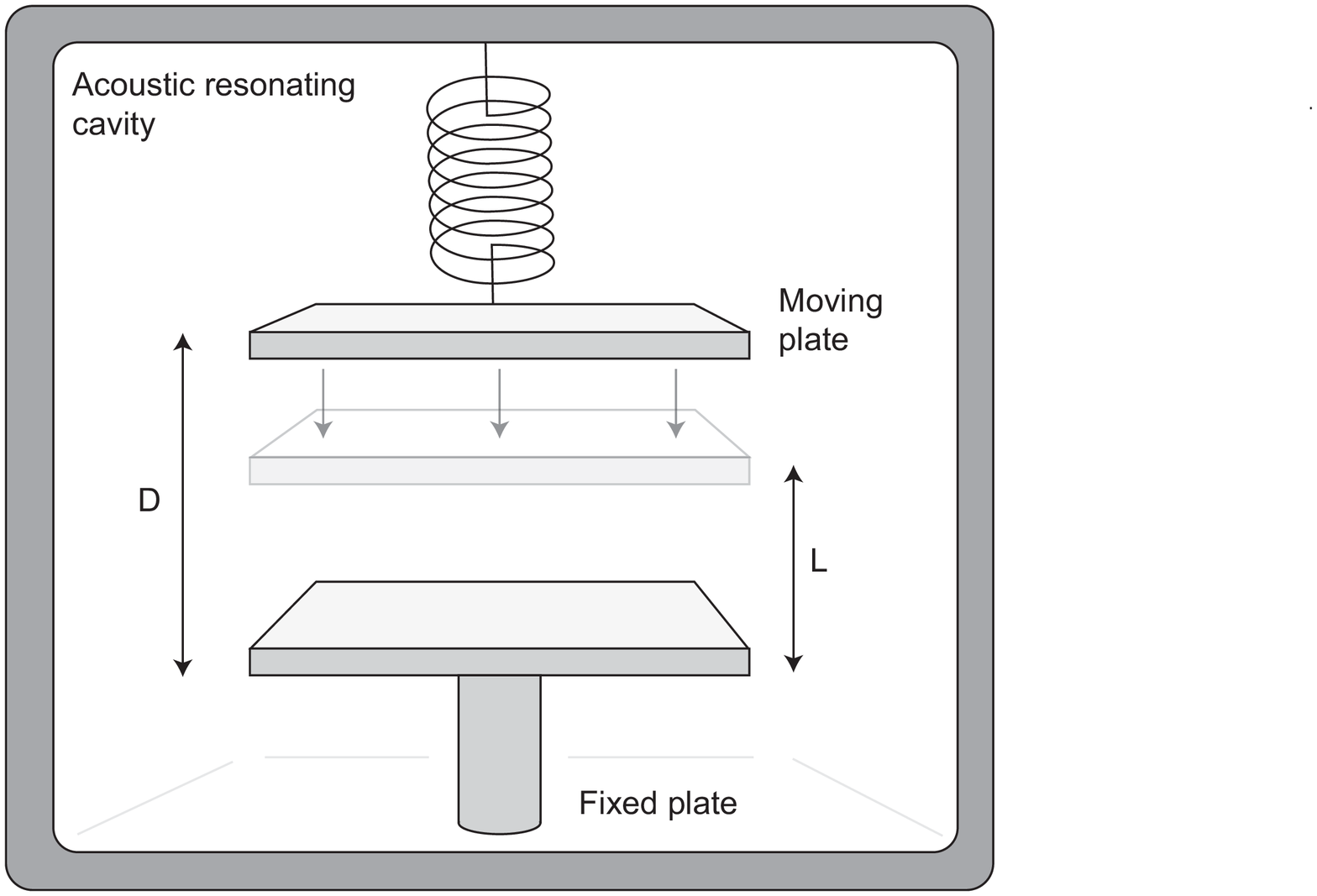}
\caption{Simple lumped one degree of freedom system considered in the calculation of the acoustic Casimir force.}
\label{fig.1}
\end{figure}

\begin{figure}
\includegraphics[width=8.cm]{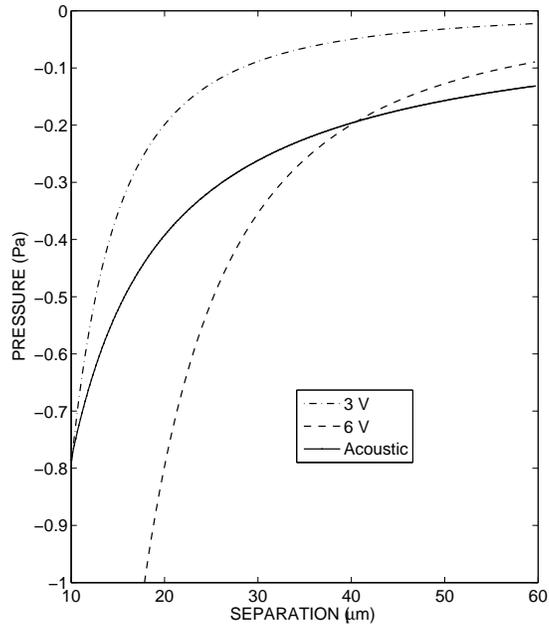}
\caption{Comparison between the acoustic Casimir pressure and the electrostatic pressure between two parallel plates, for two different values of the potential difference between the plates. The pressures are of the same order of magnitude. In this figure we use the ACP given by Eq. (\ref{lifshitz3}) which is always attractive. } 
\label{fig.1}
\end{figure}

 \begin{figure}
\includegraphics[width=8.cm]{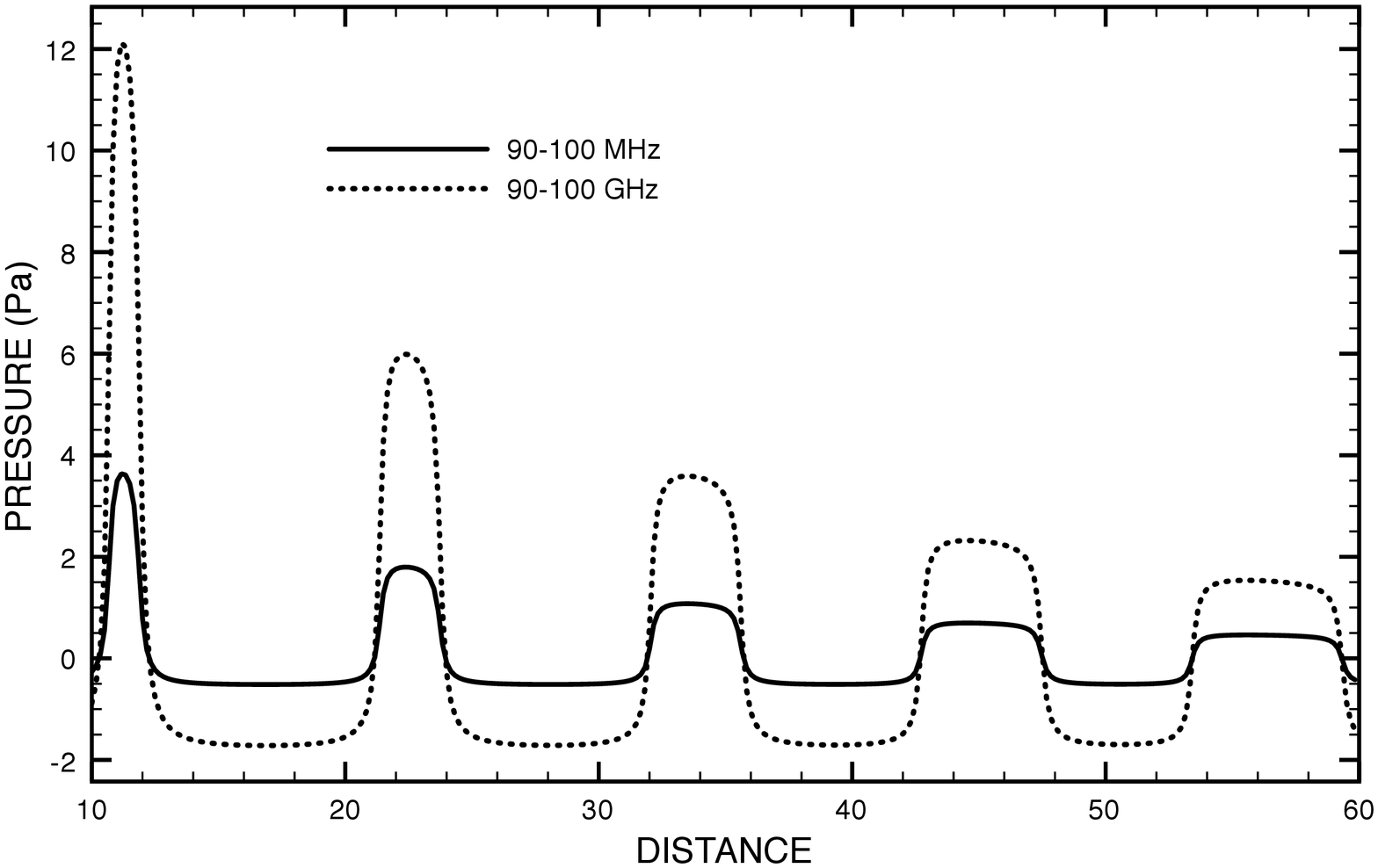}
 \caption{Acoustic Casimir Pressure as a function of separation for the frequency bandwidth $[90,100]$MHz (solid line) and $[90,100]$GHz (dotted line). The change in sign from attractive (negative) to repulsive (positive) happens when   $L\sim n c /\omega_1$.  The ACP was calculated using Eq. (\ref{lifshitz2}) and assuming an acoustic reflectivity of $r=0.8$. The horizontal axis is in microns for the mega Hertz bandwidth and nano meters for the giga Hertz bandwidth. For this figure $I_{\omega}=10^{-4}$$Watts\cdot s^{-1}\cdot m^{-2}$ for the solid line and $I_{\omega}=10^{-3}$$Watts\cdot s^{-1}\cdot m^{-2}$ for the dotted line. 
.  
 }
 \end{figure}

 \begin{figure}
\includegraphics[width=8.cm]{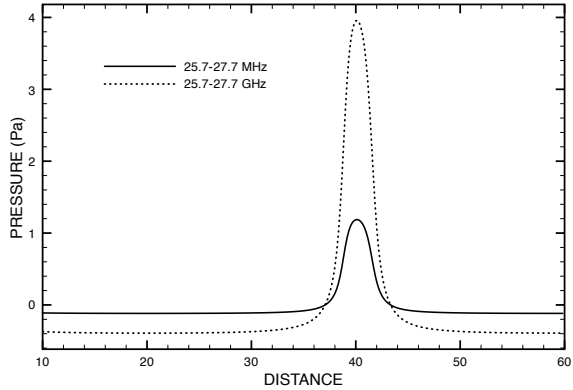}
 \caption{The separation at which  the ACP is positive can be fixed by  selecting the  frequency bandwidth
, in this case  $[25.7,27.7]$MHz(GHz) for the solid (dotted) lines, the separation at which  the ACP is positive can be fixed. In this case a sharp repulsive pressure is obtained at a separation of $L=40$$\mu m$(solid line) or $L=40$$n m$ (dotted line).  The units of the horizontal axis are the same as in Figure 3. }
 \end{figure}

\begin{figure}
\includegraphics[width=8.cm]{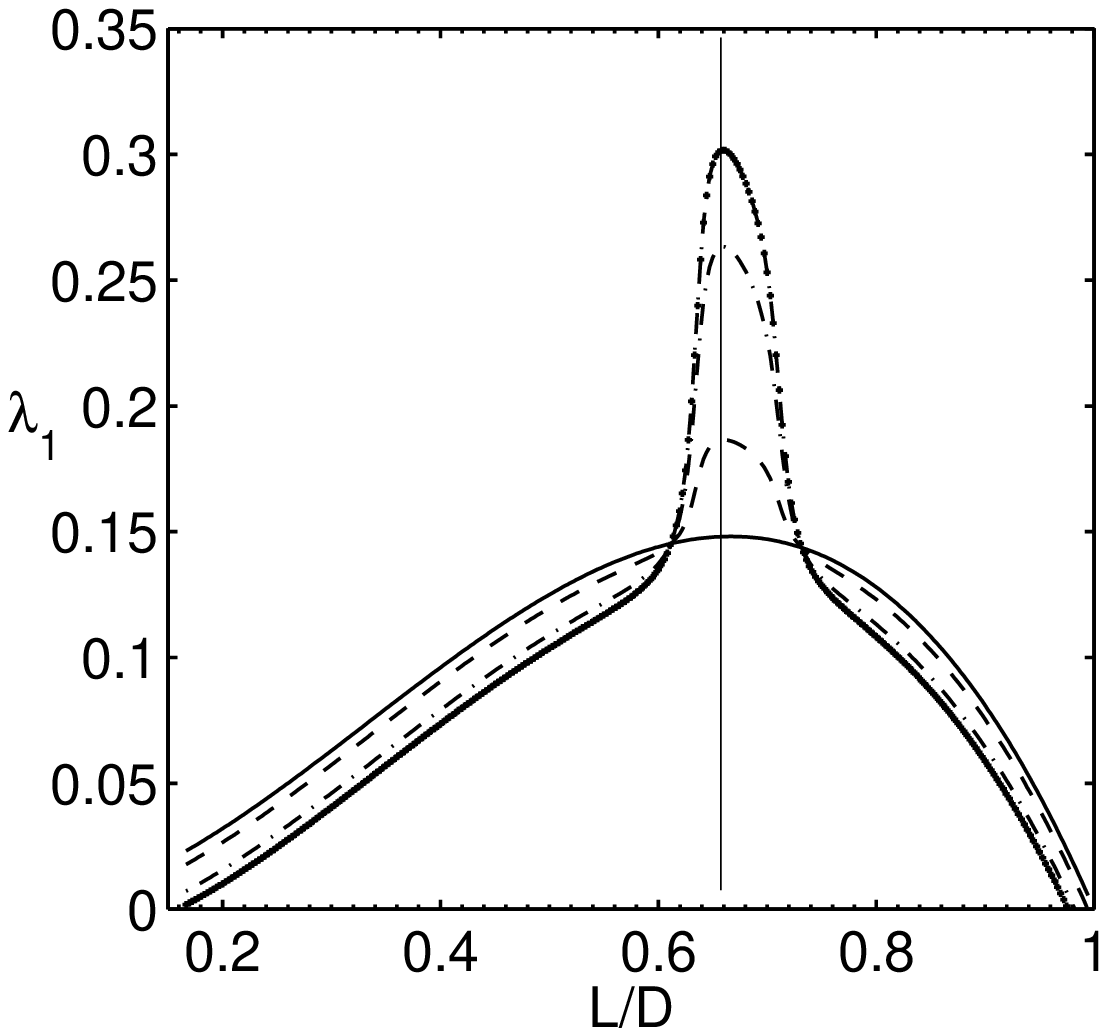}
 \caption{Bifurcation diagram for an electrostatically actuated simple-lumped system when the acoustic Casimir force is present. In particular, for two parallel plates of area $A$, we calculate the force from the ACP presented in Figure 3. In this case, the position of the maximum does not change. The different curves correspond to different values of the parameter $\lambda_2$. The vertical line is a visual aid and is located at $L/D=2/3$.  From top to bottom the different curves correspond to the values of $\lambda_2 =0.2,0.015,0.005,0$.}
 \end{figure}


\begin{thebibliography}{99}

\bibitem{casimir} H. B. G. Casimir, Proc. Kon. Ned. Akad. Wet. {\bf51} (1948) 793.

\bibitem{milonni} P. W. Milonni, ~{\it The quantum vacuum: an introduction to quantum electrodynamics}, Academic Press (1994).

\bibitem{larraza} A. Larraza, Am. J. Phys. {\bf 67}  (1999) 1028.

\bibitem{larraza2}
    A. Larraza, B. Denardo, Phys. Lett. A  {\bf 248}  ( 1998) 151.

\bibitem{villarreal} L. M. Procopio, C. Villarreal, L. W. Moch\'an, J. Phys. A {\bf 39}  (2006) 6679.

\bibitem{bschorr} O. Bschorr, J. Acoust. Soc. Am. {\bf 106} (1999) 3730.

\bibitem{george07} G. Palasantzas, J. Appl. Phys. {\bf 101} (2007) 065548.

\bibitem{footnote1} The spectral intensity $I_{\omega}$ in the frequency bandwidth $\Delta \omega$ is defined in terms of the time-averaged acoustic intensity $I$, as $I_{\omega} =I/\Delta{\omega}$.
    
\bibitem{jasa} J. B\'arcenas, L. Reyes, R. Esquivel-Sirvent, J. Acoust. Soc. Am. {\bf 116} (2004) 717.   

\bibitem{degertekin} F. L. Degertekin, B. Hadimioglu, T. Sulchek and C. F. Quate, Appl. Phys. Lett. {\bf 78} (2001) 1628.

\bibitem{huynh}  A. Huynh, N. D. Lanzillotti-Kimura, B. Jusserand,  B. Perrin, A. Fainstein, M. F. Pascual-Winter, E. Peronne and A. Lemaitre, Phys. Rev. Lett. {\bf 97} (2006) 115502.  

\bibitem{uchida} T. Uchida, A. Hamano, N. Kawashima and S. Tekeuchi, IEEE International Ultrasonic, Ferroelectrics and Frequency Control Joint 50th Anniversary Conference,  1635 (2004).

\bibitem{zhou}  Y. Zhou, L. Zhai, R. Simmons, and P. Zhonga,  J Acoust. Soc. Am.  {\bf 120} (2006) 676. 

\bibitem{batra}R. C. Batra, M. Porfiri, D. Spinello,Smart Mater. Struc. {\bf 16} (2007) 23.  

\bibitem{pelesko} J. A. Pelesko and D. H. Bernstein, {\it Modeling MEMS and NEMS} , (Chapman and Hall/CRC, Boca Raton, Florida, 2003).


\bibitem{delrio} Delrio, F. W., M. P. DeBoer, J. A. Knapp, E. D. Reedy, P. J. Clews and M. L. Dunn, Naturematerials {\bf 4} (2005) 629.

\bibitem{zhao07}  W. H. Lin and Y. P. Zhao, J. Phys. D: Appl. Phys. {\bf 40} (2007) 1649. 

\bibitem{spinello07} R. C. Batra, M. Porfiri and D. Spinello, Sensors {\bf 8} (2008)  1048; EPL {\bf 77}  (2007) 20010.

    
\bibitem{nguyen} N.-T. Nguyen, X. Huang, T. K. Chuan,  J. Fluids Eng. {\bf 124} (2002) 384. 

\bibitem{metaac} N. Fang, D. Xi, J. Xu, M. Ambati, W. Srituravanich,C. Sung, and X. Zhang, Nature Materials {\bf 5} (2006) 425.  

\end{thebibliography}
\end{document}